\documentclass[aps,prd,amsmath,amssymb,preprint,superscriptaddress,longbibliography]{revtex4-1}

\usepackage[utf8]{inputenc}
\usepackage[T1]{fontenc}
\usepackage{graphicx}
\usepackage[dvipsnames]{xcolor}
\usepackage{dcolumn}
\usepackage{bm}
\usepackage[colorlinks=true, linkcolor=RoyalBlue, citecolor=RoyalBlue]{hyperref}
\usepackage[per-mode=symbol, separate-uncertainty]{siunitx}
\usepackage[capitalise]{cleveref}
\usepackage{enumerate}
\usepackage{float}
\usepackage{lineno}
\usepackage{mhchem}

\begin{document}
\title{Role of photonic interference in exciton-mediated magneto-optic responses}

\author{Güven Budak}
\affiliation{Zentrum für QuantumEngineering (ZQE), Technical University of Munich, Garching, Germany}
\affiliation{Department of Physics, TUM School of Natural Sciences, Technical University of Munich, Munich, Germany}
\affiliation{Munich Center for Quantum Science and Technology (MCQST), Technical University of Munich, Garching, Germany}

\author{Christian Riedel}
\affiliation{Zentrum für QuantumEngineering (ZQE), Technical University of Munich, Garching, Germany}
\affiliation{Department of Physics, TUM School of Natural Sciences, Technical University of Munich, Munich, Germany}

\author{Akashdeep Kamra}
\affiliation{Department of Physics and Research Center OPTIMAS, Rheinland-Pfälzische Technische Universität Kaiserslautern-Landau, Kaiserslautern, Germany}

\author{Patrick Rinke}
\affiliation{Department of Physics, TUM School of Natural Sciences, Technical University of Munich, Munich, Germany}
\affiliation{Atomistic Modeling Center, Munich Data Science Institute, Technical University of Munich, 85748 Garching, Germany}
\affiliation{Munich Center for Machine Learning (MCML), 85748 Garching, Germany}

\author{Christian Back}
\affiliation{Zentrum für QuantumEngineering (ZQE), Technical University of Munich, Garching, Germany}
\affiliation{Department of Physics, TUM School of Natural Sciences, Technical University of Munich, Munich, Germany}
\affiliation{Munich Center for Quantum Science and Technology (MCQST), Technical University of Munich, Garching, Germany}

\author{Matthias Stosiek}
\email{matthias.stosiek@tum.de}
\affiliation{Department of Physics, TUM School of Natural Sciences, Technical University of Munich, Munich, Germany}
\affiliation{Atomistic Modeling Center, Munich Data Science Institute, Technical University of Munich, 85748 Garching, Germany}
\affiliation{Munich Center for Machine Learning (MCML), 85748 Garching, Germany}

\author{Florian Dirnberger}
\email{f.dirnberger@tum.de}
\affiliation{Zentrum für QuantumEngineering (ZQE), Technical University of Munich, Garching, Germany}
\affiliation{Department of Physics, TUM School of Natural Sciences, Technical University of Munich, Munich, Germany}
\affiliation{Munich Center for Quantum Science and Technology (MCQST), Technical University of Munich, Garching, Germany}
	
%\date{\today}
\begin{abstract}
Coupled optical and magnetic excitations can give rise to remarkably strong magneto-optic responses. 
This is particularly evident in van der Waals magnets, such as the antiferromagnet CrSBr, where excitons and magnons emerge from the same electronic orbitals. 
While previous work has primarily focused on uncovering the magneto-electric origin of the resulting exciton–magnon interactions, the influence of photonic effects has received comparatively little attention.
Here, we use numerical simulations to disentangle exciton–magnon coupling from the exciton-mediated magnon-photon interactions observed in optical experiments. 
Our simulations show the strong dependence of these interactions on photonic interference and dispersion effects near excitonic resonances. 
Such effects shape the optical response to coherent magnons and make it intrinsically non-linear in the magnon-induced exciton energy shift.
Thermal magnons, which have a particularly pronounced impact on excitons, are found to even produce qualitatively different trends in optical signatures. 
Depending on weak or strong coupling of excitons and photons, the same exciton–magnon interaction can lead to a red-shift of optical modes, a nearly vanishing response, or their blue-shift.
Finally, we demonstrate first steps towards optimizing the multi-parameter problem of efficient magnon–photon transduction using a machine-learning approach.
\end{abstract}

\maketitle

%\begin{linenumbers}		
\section{Introduction}

Light–matter interactions are often strongly enhanced in materials that host narrow and intense excitonic resonances. When these excitons arise in magnetic materials, their large oscillator strength can in principle amplify magneto–optic effects and provide sensitive optical access to spin dynamics. Recent studies on van der Waals (vdW) magnets have shown that this expectation is well founded: layered magnetic crystals such as CrSBr exhibit unusually strong optical responses to changes in their spin configuration~\cite{wilson2021interlayer,bae2022exciton,dirnberger2023magneto}, suggesting that excitons mediate an efficient coupling between magnons and the photon field.

In conventional magneto–optic materials, phenomena such as the magneto-optic Kerr or Faraday effect originate from small magnetization–dependent corrections to a mostly smooth and weakly dispersive dielectric function~\cite{weiglhofer2003introduction}. 
In this regime, photonic thin-film interference often influences only the magnitude of the optical response, but its sign and spectral structure remain determined by magnetization~\cite{visnovsky2018optics}. 
Excitonic vdW magnets, however, typically operate in a different regime. 
In such materials, excitons dominate the dielectric function and strongly modify its dispersion. 
As a consequence, even modest exciton–magnon interactions can induce large and spectrally structured changes that are further shaped by interference in realistic multilayer geometries.
Magneto–optic responses therefore not only reflect the underlying spin physics but also the photonic environment.

A prototypical example of this is the layered antiferromagnet CrSBr~\cite{telford2020layered}.
In this material, tightly bound excitons~\cite{liebich2025controlling} with large oscillator strength~\cite{dirnberger2023magneto} give rise to isolated, narrow optical resonances in the near-infrared and optical domain~\cite{wilson2021interlayer, klein2023bulk,smiertka2026distinct, meineke2024ultrafast}. 
As discussed in more detail below, excitons and magnons in CrSBr are intimately coupled with a multitude of consequences for its optical properties. 
The large oscillator strength of CrSBr excitons, together with the ease of integrating vdW materials into microcavities, has recently also enabled studies of magneto-optics in the strong exciton-photon coupling regime. 
Initial work highlighted the influence of exciton–photon hybridization~\cite{dirnberger2023magneto}, photonic anisotropy~\cite{ruta2023hyperbolic, li2024two}, and magnon-induced exciton nonlinearities~\cite{datta2025magnon}. 
Very recent studies have even begun to probe magneto–optic properties of exciton-polariton condensates~\cite{zhang2025exciton, han2025exciton}. 
So far, however, most efforts have focused on identifying the microscopic magnetoelectric origin of exciton–magnon coupling, leaving the role of photonic interferences~\cite{moon1993interference, pasanen2024transient} largely unaddressed.

In this work, we investigate how photonic interference and dispersion influence the magneto–optical response of CrSBr. 
Using numerical simulations, we reproduce and interpret features observed in recent optical experiments and show that interference in realistic multilayer structures can substantially reshape the apparent signatures of excitons coupled to coherent magnons. 
We further analyze the impact of thermal magnons on the excitonic dielectric function and demonstrate that their contribution can produce qualitatively different optical outcomes in weakly and strongly coupled exciton–photon regimes. 
Finally, we illustrate that machine–learning–assisted optimization provides a practical strategy for identifying photonic geometries that maximize the exciton-mediated magnon-photon coupling for transduction. 

\section{Exciton-magnon interactions}
\begin{figure*}[]
	\includegraphics[]{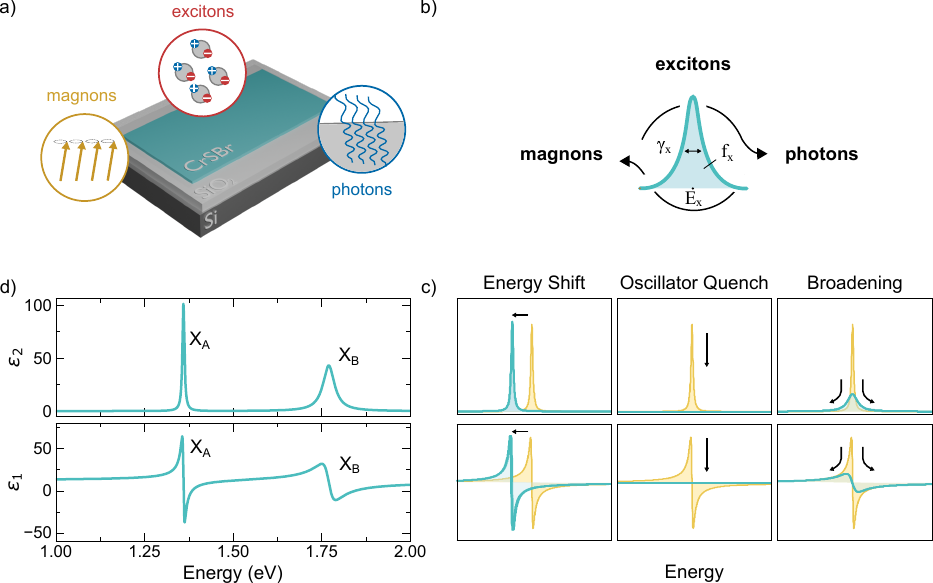}
	\caption{\textbf{Role of magnons in the exciton-dominated range of the dielectric function.} 
		a)~Sketch of a thin CrSBr crystal on top of a silicon substrate covered by an oxide layer. 
		b)~Magnons couple to photons by changing the exciton resonance.
        c)~Exciton-magnon interactions can shift, quench, or broaden the exciton resonance.
		d)~Dielectric function of CrSBr in the near infrared region modeled using parameters from Ref.~\cite{shao2025magnetically}.
		 }
	\label{fig:1}
\end{figure*}

To understand the impact of magnetization on optical properties, it can be insightful to consider that magnetic insulators and semiconductors may host a zoo of mutually interacting quasiparticles (\cref{fig:1}a), b)).
Magnetic materials studied in the context of magnons are often weakly dispersing in the optical frequency domain (cf. Fig. S1). 
However, when excitons dominate the optical spectrum, different mechanisms for the coupling of magnons to light become possible. 

To illustrate these mechanisms, it is instructive to consider a model dielectric function $\varepsilon(E)$ at energies $E$ close to an exciton resonance (\cref{fig:1}b)), while all other contributions are included in a background relative permittivity $\varepsilon^\infty$:
\begin{align}
	\varepsilon(E) = \varepsilon^\infty - \frac{f_{X}}{E^2 - E_X^2 + i\gamma_X E}. \label{Eq:1}
\end{align}
In this model, magnons can couple to the photon field by shifting the exciton energy $E_X$, quenching the oscillator strength $f_X$, or inducing  broadening of the exciton line-width $\gamma_X$.
The impact of these processes on the dielectric function is schematically depicted in \cref{fig:1}c). 

For light polarized along the $b$--axis, the dielectric function of CrSBr can be approximated by this model. 
\Cref{fig:1}d) shows a plot of its low-temperature dielectric function $\epsilon_b(\omega)$ in the vicinity of both $A$ and $B$ exciton resonances. 
It is derived from \cref{Eq:1} using parameters determined in recent experiments~\cite{shao2025magnetically}.
Note that surface excitons are omitted in our analysis since their effects are small for crystals with more than a few layers.
In the following, we use this dielectric function to numerically calculate reflectance spectra of realistic multilayer sample structures via the transfer-matrix method (see Supplementary Section S2).
This allows us to analyze the role of photonic interferences in the different optical experiments demonstrating exciton-magnon coupling.
Unless specified otherwise, all spectra below are calculated for a lattice temperature of 20\,K, samples with infinite lateral extensions, and light polarized along the crystallographic $b$--axis.

\section{Coherent magnons in a canted spin state}

\begin{figure*}[t]
	\centering
	\includegraphics[]{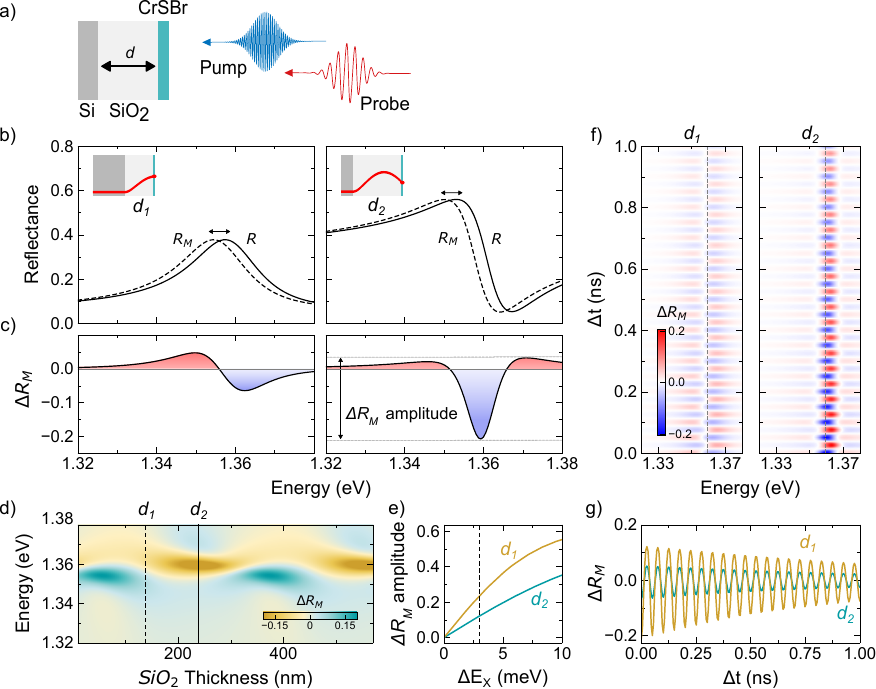}
	\caption{\textbf{Coherent magnon-induced reflectance modulation mediated by excitons.} 
		a)~Schematic pump-probe measurement of a 5\,nm-thin CrSBr crystal on top of a Silicon substrate covered by an oxide of thickness $d$.
        b)~Calculated reflectance spectra with ($R_M$) and without ($R$) a shift of the exciton energy induced by magnons. Left and right panels compare spectra obtained at oxide thickness $d_1 = 140$\,nm and $d_2=240$\,nm. The electric field distribution calculated at $E = E_X$ respectively exhibits a maximum or minimum at the position of the CrSBr crystal for these thicknesses, as indicated in the inset. 
        c)~Magnon-induced reflectance contrast, $\Delta R_M  = R_M - R$, and amplitude $|\Delta R_{M,max}  - \Delta R_{M,min}|$.
        d)~$\Delta R_M$ as a function of energy and \ce{SiO2} thickness. 
        e)~Maximum amplitude of $\Delta R_M$ as a function of the magnon-induced exciton energy shift $\Delta E_X$ for $d=d_1$ and $d=d_2$. 
        Dashed line indicates the exciton energy shift used in our simulations.
		f)~Energy and time-dependence of $\Delta R_M$. 
		g)~Line cuts from f) taken at the maximum amplitude of $\Delta R_M$ for $\Delta t = 0$.}
	\label{fig:2}
\end{figure*}

The most unambiguous optical signature of magnons has been observed in pump-probe measurements~\cite{bae2022exciton, dirnberger2023magneto, diederich2023tunable, sun2024dipolar}. 
In a seminal experiment, Bae \textit{et al.} showed that ultrafast optical pump pulses used to create coherent collective oscillations of spins in a canted state cause a periodic modulation of the interlayer spin angle that can be detected by changes in the reflection of time-delayed probe pulses~\cite{bae2022exciton}.
This experiment is sketched in \cref{fig:2}a).

The underlying coupling of excitons to coherent magnons can be modeled by imposing a damped oscillation of the exciton energy 
\begin{equation}
\Delta E_X(t) = \Delta E_{X}(0) \cos(\omega_M t) \exp(-t/\tau_M)
\end{equation} 
in the dielectric function of \cref{Eq:1}.
For our analysis, we use $\Delta E_{X}(0) =3$\,meV and $\omega_M = 20$ GHz. We also introduce a phenomenological magnon decay characterized by a time constant $\tau_M = 1$\,ns, in agreement with experimental observations~\cite{bae2022exciton}. 
Our model thus implicitly assumes both uniform spatial occupation of magnons within the probe volume and the fact that oscillator strength and line-width of the exciton resonance are not strongly modified by coherent magnons. 
The latter, in particular, is supported by static reflectance measurements~\cite{wilson2021interlayer, klein2023bulk} indicating that the primary effect of modifying the interlayer spin angle is a change in the exciton resonance energy $E_X$. 

To illustrate the role of photonic interferences in the optical response to coherent magnons, we exemplarily calculate the reflectance of a 5\,nm-thin crystal on top of an oxide-covered silicon substrate. 
Since the thin CrSBr crystal itself does not provide enough optical confinement for self-hybridized polaritons~\cite{dirnberger2023magneto}, excitons remain weakly coupled to photons. 
Nonetheless, optical modes are still weakly confined in the oxide layer beneath the CrSBr crystal, as indicated in the insets of \cref{fig:2}b).
These modes are subject to constructive or destructive interference depending on oxide layer thickness. 
Reflectance in the vicinity of the exciton resonance thus varies strongly with oxide thickness (\cref{fig:2}d)).  

Our discussion focuses on the two cases of constructive ($d_{1}=140$\,nm) and destructive ($d_{2}=240$\,nm) interference where excitonic reflectance is minimized or maximized, respectively. 
\Cref{fig:2}b), c) demonstrates the changes in the magnon-induced reflectance contrast, $\Delta R_M = R - R_M$, that occur between them. 
We find that not only the magnitude, but also the spectral shape of $\Delta R_M$ is altered by photonic interferences in the \ce{SiO2} layer.
Comparing the two cases, $d_1$ and $d_2$, therefore illustrates the difference between an exciton-mediated magneto-optic responses measured, e.g., in pump-probe experiments, and the underlying exciton-magnon coupling.
\Cref{fig:2}e) shows another important implication of the fact that the optical response to coherent magnons is mediated by excitons.
The response is inherently non-linear in the magnon-induced exciton shift $\Delta E_{X}(0)$ due to the non-linear optical dispersion near exciton resonances. 
Like the spectral shape of the resonances, this non-linearity depends on oxide thickness. 
It is, however, only pronounced at large $\Delta E_{X}(0)$.

\Cref{fig:2}f) shows the corresponding time- and energy-resolved pump-probe response of $\Delta R_M$.  
At destructively interfering optical fields, for $d = d_2$, the amplitude of $\Delta R_{M}$ is more than two times larger.
If highly reflective layers such as gold films are part of the stack, this difference even exceeds one order of magnitude (cf. Fig. S2).
In the spectral vicinity of exciton resonances, photonic interferences thus play a significant role for exciton-mediated optical responses to coherent magnons. 

\section{Thermal magnons and spin disorder}
\begin{figure*}[t]
	\centering
	\includegraphics[]{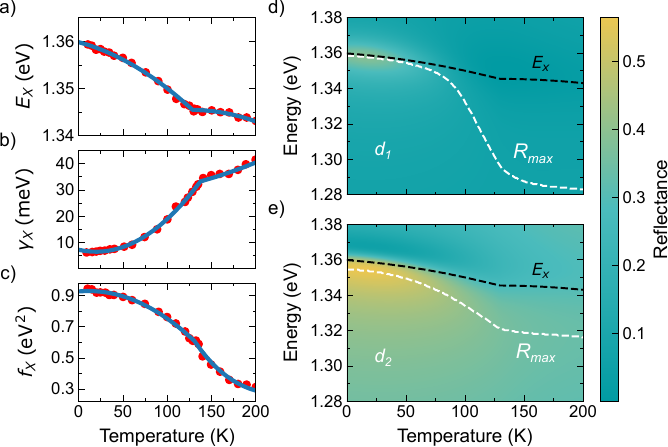}
	\caption{\textbf{Temperature dependence of the optical response.} 
		a-c)~Temperature dependence of excitonic parameters determined by the experiments in Ref.~\cite{shao2025magnetically} (red dots) approximated by polynomial functions (blue lines) used for calculations. 
		Reflectance of a 5 nm-thin CrSBr crystal placed on a substrate with d) $d_1 = 140$\,nm and e) $d_2 = 240$\,nm oxide thickness. Black dashed lines show $E_X(T)$}
	\label{fig:3}
\end{figure*}

Magnetic disorder induced by increasing lattice temperature leads to the excitation of incoherent thermal magnons. These magnons randomize the spin alignment not only between adjacent layers but also within individual layers. 
As a consequence, several parameters of the exciton resonance are modified simultaneously. 
Experiments indicate that the exciton energy $E_X$, oscillator strength $f_X$, and linewidth $\gamma_X$ all exhibit a pronounced temperature dependence~\cite{dirnberger2023magneto, shao2025magnetically, datta2025magnon}. 
The experimentally extracted values~\cite{shao2025magnetically} used in our simulations are shown in \cref{fig:3}a)--c).
Polynomial fits serve as phenomenological parameterizations. 

Several microscopic mechanisms contribute to these trends. 
Analytical models that incorporate interlayer spin disorder predict a thermal shift of the exciton energy that scales with the density of incoherent magnons in each magnetic sublattice~\cite{dirnberger2023magneto}. 
This mechanism therefore adds a magnetic contribution to the temperature dependence of the exciton resonance beyond the conventional lattice-induced bandgap renormalization. 
In addition, recent studies report an anomalous temperature dependence of the exciton oscillator strength $f_X$~\cite{ruta2023hyperbolic, shao2025magnetically, wang2023magnetically}. 
\textit{Ab initio} calculations attribute this effect to reduced hopping amplitudes of electrons and holes along increasingly disordered spin configurations within the layers~\cite{shao2025magnetically}. 
Finally, scattering of excitons by magnons leads to an increase of the exciton linewidth $\gamma_X$, consistent with observations from reflectance spectroscopy~\cite{shao2025magnetically}, exciton transport measurements~\cite{dirnberger2025exciton}, and angle-resolved photoemission experiments~\cite{bianchi2023paramagnetic,wu2025mott}.

Because excitonic resonances dominate the dielectric function of CrSBr in the near-infrared region, these temperature-dependent changes strongly affect the optical response. 
To illustrate this effect, we calculate the reflectance of a 5\,nm-thin CrSBr crystal placed on a \ce{Si}/\ce{SiO2} substrate for two representative oxide thicknesses corresponding to constructive and destructive interference. 
The resulting reflectance maps are shown in \cref{fig:3}d),~e).

The comparison reveals that the experimentally observable reflectance maxima $R_{\text{max}}(T)$ deviate substantially from the intrinsic exciton energy $E_X(T)$. 
This discrepancy arises because the simultaneous modification of oscillator strength and linewidth alters the interference conditions of the optical field. 
The effect is particularly pronounced in the regime of constructive interference. 
Consequently, extracting intrinsic excitonic parameters from reflectance measurements requires quantitative modeling that explicitly accounts for photonic interference effects.

\section{Magnons in the regime of strong exciton-photon coupling}
Recent experiments have demonstrated strong coupling between excitons and photons in CrSBr, resulting in exciton-polariton formation both in intrinsic crystals and in external microcavities~\cite{dirnberger2023magneto, wang2023magnetically}, on photonic lattices~\cite{li2024two}, and in conventional microcavities~\cite{zhang2025exciton}. 
In the following, we investigate how this regime modifies magnon-induced optical responses using transfer-matrix calculations. 
Our analysis focuses on a prototypical planar microcavity in which a 5\,nm-thick CrSBr crystal is embedded in a $\lambda/4$ cavity formed by two planar mirrors (cf. \cref{fig:4}a). 
The bottom gold mirror is highly reflective (infinite thickness), while the 40\,nm-thin top gold mirror is semi-transparent. 
The angle-dependent reflectance spectrum of this microcavity exhibits a lower and an upper polariton branch with varying exciton $\boldmath |X(\theta)|^2$ and photon $\boldmath |P(\theta)|^2$ fractions. 
These fractions can be described using a standard Hopfield model~\cite{deng2010exciton} and tuned experimentally via the spacer thickness (cf. \cref{fig:4}b),d) inset). 
As in most experiments~\cite{dirnberger2023magneto,wang2023magnetically}, we focus on the reflectance response of the lower polariton branch at normal incidence.

\subsection{Coherent magnons}
\begin{figure*}[t]
    \centering
	\includegraphics[]{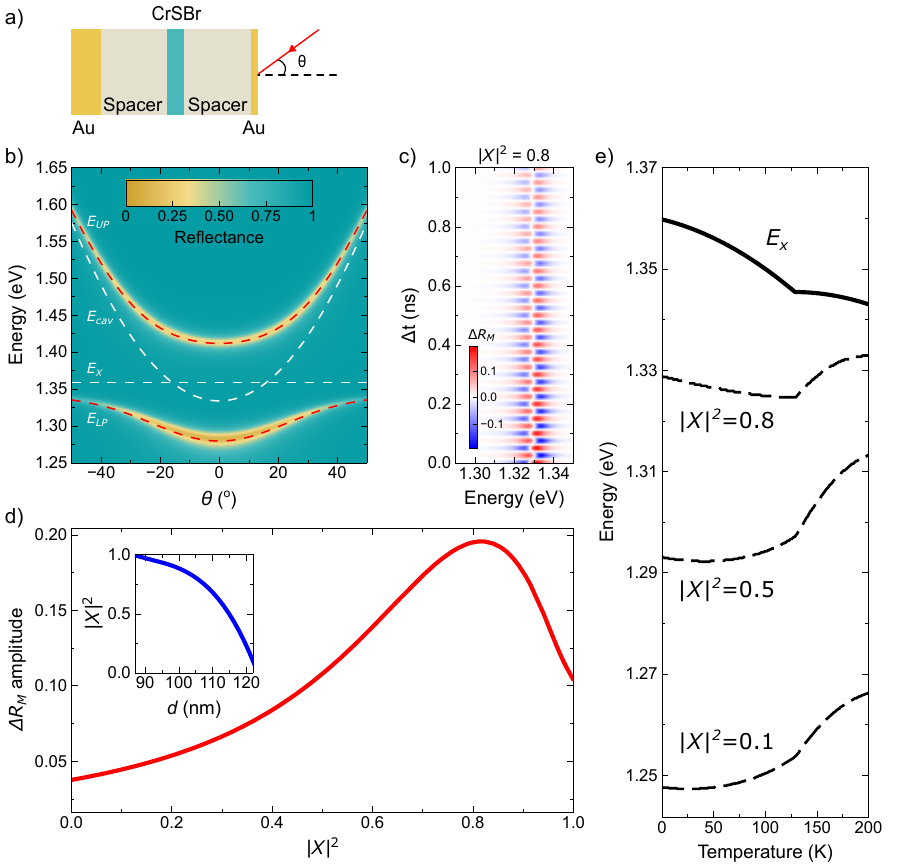}
	\caption{\textbf{Magneto-optic responses of a CrSBr microcavity.} 
	a)~Microcavity structure.
	b)~Reflectance calculated for spacer thickness $d =$ 117\,nm at different angle of incidence, $\theta$. 
    Fits of upper ($E_{UP}$) and lower ($E_{LP}$) polariton energies derived from a Hopfield model~\cite{deng2010exciton} are plotted as red dashed curves. 
    Energy of bare exciton ($E_X$) and cavity mode ($E_{cav}$) are indicated by white dashed lines.
	c)~Transient reflectance induced by coherent magnons for $|X(0)|^2=0.8$.
	d)~Amplitude of $\Delta R_M$ as a function of exciton fraction. Inset: Dependence of $|X|^2$ on spacer layer thickness for $\theta = 0^\text{o} $.
	e)~Temperature dependence of lower polariton energies for different exciton fractions $|X(0)|^2$. 
    }
\label{fig:4}
\end{figure*}

To model coherent magnons in the strong-coupling regime, we impose the same 3\,meV exciton-energy shift as in Sec.~III. 
The calculated reflectance response is shown in \cref{fig:4}c), d). 
Interestingly, the amplitude of the magnon-induced reflectance modulation $\Delta R_M$ does not peak for purely excitonic polaritons. 
Instead, the largest response occurs at an exciton fraction of approximately $|X|^2 \approx 0.8$.

This behavior reflects the hybrid nature of polaritons. 
The magnon-induced shift originates from the excitonic component, while the photonic component narrows the polariton linewidth and enhances the optical visibility of the resonance shift. 
As a result, the strongest signal is obtained for polaritons with mixed exciton-photon character rather than for purely excitonic states.

\subsection{Thermal magnons}
The microcavity further enhances the impact of temperature on the optical spectrum of CrSBr. 
To illustrate this, \cref{fig:4}e) shows the calculated temperature dependence of the lower-polariton energy for different exciton fractions $|X(0)|^2$. 
Polaritons with large excitonic content follow the intrinsic red-shift of the exciton energy $E_X(T)$ discussed in Sec.~IV (cf. \cref{fig:3}). 
Polaritons with a larger photonic fraction, however, display qualitatively different behavior. For $|X(0)|^2 < 0.7$, the lower-polariton resonance instead shifts towards higher energies with increasing temperature.

This apparently anomalous blue-shift arises from two competing effects. 
Thermal spin disorder reduces the exciton energy, which by itself would red-shift the polariton.
At the same time, the oscillator strength decreases with temperature, reducing the Rabi splitting of the exciton-photon system. 
For lower polaritons, this second effect increases the mode energy and can dominate when the photonic fraction is sufficiently large. 
Similar behavior has already been reported in several recent experiments~\cite{dirnberger2023magneto, wang2023magnetically, han2025exciton, zhang2025exciton}.

\section{Machine learning–based optimization of thin-film structures}

Finally, we address the optimization of optical structures that maximize the magnon-induced reflectance modulation $\Delta R_M$, a quantity directly relevant for proposed magnon-photon transduction schemes~\cite{bae2022exciton,tang2025coherent}. 
In realistic device architectures, transduction would likely proceed through a sequence in which microwave photons excite magnons, magnons interact with excitons, and excitons subsequently couple to near-infrared or optical photons. 
Efficient conversion therefore requires simultaneous optimization of both microwave and optical field distributions. 
As a first step, we focus here on maximizing the optical response $\Delta R_M$.

\begin{figure*}[!ht]
    \centering
	\includegraphics[]{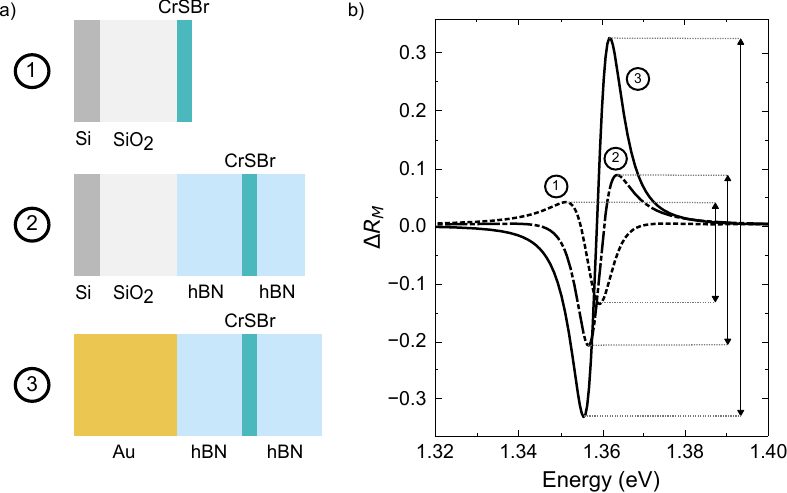}
	\caption{\textbf{Optimization of multilayer stacks.} 
	a)~Schematic of different sample configurations with layer thicknesses optimized by BOSS.
    b)~Optimized $\Delta R_M$ spectra for scenarios (1) to (3). Vertical arrows indicate the amplitude of each $\Delta R_M$ curve.}
\label{fig:5}
\end{figure*}	

Even for a simple CrSBr--\ce{SiO2}--Si stack, $\Delta R_M$ exhibits a strongly periodic dependence on the \ce{SiO2} thickness due to interference, leading to multiple local extrema (cf. \cref{fig:2}d)). 
For more complex multilayer geometries, identifying the global optimum becomes a high-dimensional optimization problem. We therefore employ the Bayesian Optimization Structure Search (BOSS) algorithm~\cite{boss}, which efficiently explores the parameter space by iteratively constructing a probabilistic surrogate model of the objective function. 
New sampling points are then selected based on this model, allowing rapid convergence towards optimal layer configurations. 
Further details of the optimization procedure, including parameter ranges and constraints, are provided in Supplementary Section~2 and shown in Fig.~S3.

We start from a standard structure consisting of a 5\,nm-thin CrSBr flake on a silicon substrate with 285\,nm of \ce{SiO2}. 
For this stack, the modulation amplitude is approximately $\Delta R_M \approx 0.2$. 
The optimization indicates that encapsulating the CrSBr layer between two hBN layers already increases this value to roughly 0.3. Introducing a metallic mirror below the structure further enhances the signal and yields amplitudes exceeding 0.7. 
\Cref{fig:5} shows the optimized layer stacks and the corresponding $\Delta R_M$ spectra.

Our simulations further indicate that improvements in crystal quality could significantly enhance the achievable response. 
In particular, reducing the exciton linewidth to $\gamma_X \approx 1$\,meV would allow the reflectance modulation to approach its theoretical maximum value of $\Delta R_M \approx 2$.

\section{Conclusion}

Excitons in magnetic insulators can mediate a pronounced coupling between magnons and photons, but our analysis shows that the resulting optical signatures depend sensitively on the photonic environment. 
Unlike in conventional magneto–optics, where the dielectric response is often weakly dispersive, excitonic vdW magnets such as CrSBr operate in a regime where interference, dispersion, and magnetic effects are strongly intertwined. 
Photonic interference can therefore amplify, suppress, or even invert spectral shifts induced by magnons.
Using transfer-matrix simulations, our study finds that this interplay is relevant in both weak and strong exciton–photon coupling regimes and that thermal magnons and spin fluctuations can produce opposite energy shifts depending on the polariton composition. 
Careful engineering of the photonic environment, combined with machine-learning–based optimization, can substantially enhance the sensitivity of optical measurements to magnons, which may prove valuable for future schemes aiming to transduce magnetic excitations into optical signals in quantum technologies.
%\end{linenumbers}
	
%%%%%%%%%%%%%%%%%%%%%%%%%%%%%%%%%%%%%%%%%%%%%%%%%%%%%%%%%%%%%%%%%%%%%%%%
%
%References: 
%
%%%%%%%%%%%%%%%%%%%%%%%%%%%%%%%%%%%%%%%%%%%%%%%%%%%%%%%%%%%%%%%%%%%%%%%%

\clearpage

\bibliography{Reflectance_paper}
\bibliographystyle{naturemag}

%\section{Methods}
%\subsection{Optical spectroscopy and time-resolved microscopy}

\noindent\textbf{Acknowledgments:}
We thank Prof. Alexey Chernikov and Prof. Geoffrey Diederich for valuable discussions. 
F.D. and G.B. gratefully acknowledge financial support by the DFG via the Emmy Noether Program (Project-ID 534078167) and the Munich Center for Quantum Science and Technology (MCQST) under Germany’s Excellence Strategy – EXC-2111 - 390814868. 
A.K. acknowledges financial support from the German Research Foundation (DFG) via Spin+X TRR 173-268565370 (project A13).
The work of PR and MS is funded by the Deutsche Forschungsgemeinschaft (DFG, German Research Foundation) under Germany’s Excellence Strategy – EXC 2089/2 – 390776260.

%\noindent\textbf{Author contributions:}
\end{document}

\title{SUPPLEMENTARY INFORMATION:\\Role of photonic interference in exciton-mediated magneto-optic responses} 

\author{Güven Budak}
\affiliation{Zentrum für QuantumEngineering (ZQE), Technical University of Munich, Garching, Germany}
\affiliation{Department of Physics, TUM School of Natural Sciences, Technical University of Munich, Munich, Germany}
\affiliation{Munich Center for Quantum Science and Technology (MCQST), Technical University of Munich, Garching, Germany}

\author{Christian Riedel}
\affiliation{Zentrum für QuantumEngineering (ZQE), Technical University of Munich, Garching, Germany}
\affiliation{Department of Physics, TUM School of Natural Sciences, Technical University of Munich, Munich, Germany}

\author{Akashdeep Kamra}
\affiliation{Department of Physics and Research Center OPTIMAS, Rheinland-Pfälzische Technische Universität Kaiserslautern-Landau, Kaiserslautern, Germany}

\author{Patrick Rinke}
\affiliation{Department of Physics, TUM School of Natural Sciences, Technical University of Munich, Munich, Germany}
\affiliation{Atomistic Modeling Center, Munich Data Science Institute, Technical University of Munich, 85748 Garching, Germany}
\affiliation{Munich Center for Machine Learning (MCML), 85748 Garching, Germany}

\author{Christian Back}
\affiliation{Zentrum für QuantumEngineering (ZQE), Technical University of Munich, Garching, Germany}
\affiliation{Department of Physics, TUM School of Natural Sciences, Technical University of Munich, Munich, Germany}
\affiliation{Munich Center for Quantum Science and Technology (MCQST), Technical University of Munich, Garching, Germany}

\author{Matthias Stosiek}
\email{matthias.stosiek@tum.de}
\affiliation{Department of Physics, TUM School of Natural Sciences, Technical University of Munich, Munich, Germany}
\affiliation{Atomistic Modeling Center, Munich Data Science Institute, Technical University of Munich, 85748 Garching, Germany}
\affiliation{Munich Center for Machine Learning (MCML), 85748 Garching, Germany}

\author{Florian Dirnberger}
\email{f.dirnberger@tum.de}
\affiliation{Zentrum für QuantumEngineering (ZQE), Technical University of Munich, Garching, Germany}
\affiliation{Department of Physics, TUM School of Natural Sciences, Technical University of Munich, Munich, Germany}
\affiliation{Munich Center for Quantum Science and Technology (MCQST), Technical University of Munich, Garching, Germany}

\maketitle
\tableofcontents
\clearpage

\section{Supplementary figures}

\begin{figure*}[h]
	\includegraphics[scale=0.9]{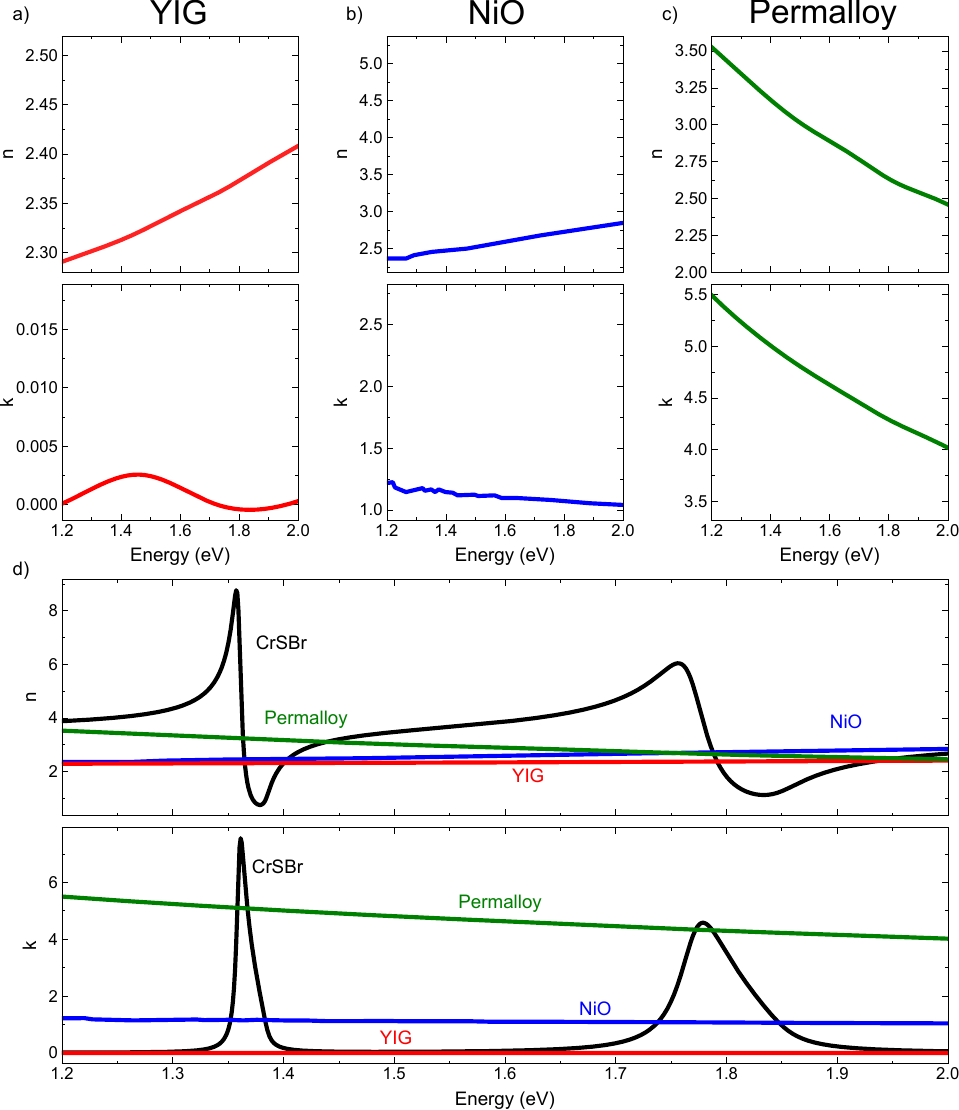}
	\caption{\textbf{Optical dispersion of different materials.} 
		Energy dependence of refractive index $n$ and extinction coefficient $k$ of a)~Yttrium Iron Garnet (YIG)~\cite{wemple1974optical}, b)~Nickel Oxide (NiO)~\cite{ahmed2021influence}, c)~Permalloy~\cite{tikuivsis2017optical}.
		d)~Comparison with highly dispersive CrSBr.}
	\label{fig:Fig-S1}
\end{figure*}	

\begin{figure*}[]
	\includegraphics[]{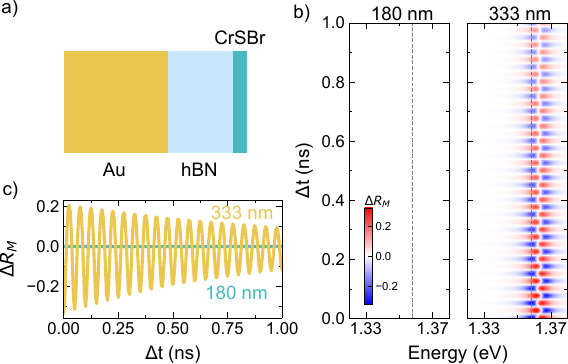}
	\caption{\textbf{CrSBr on top of a gold mirror.}
		a)~Simulated sample structure. 
		b)~Time-resolved maps of reflectance response $\Delta R_M$ for hBN-thickness of 180\,nm and 330\,nm.
		c)~Line-cuts from b) taken at the positions indicated by the dashed lines.}
	\label{fig:Fig-S2}
\end{figure*}	

\begin{figure*}[]
	\includegraphics[]{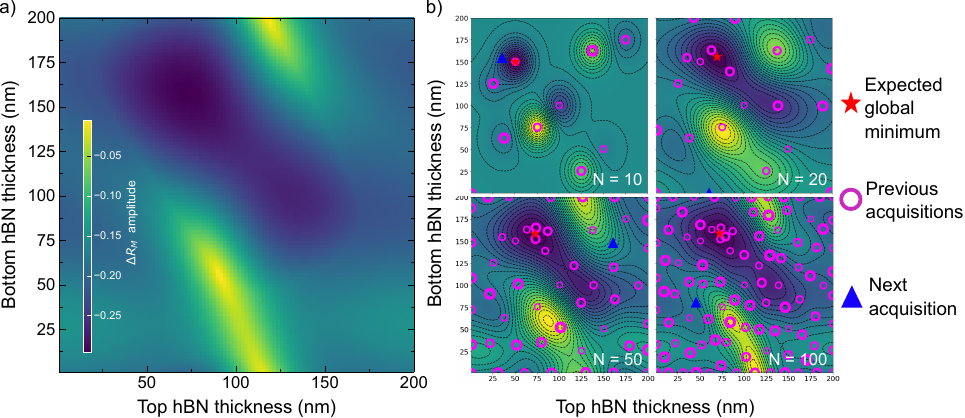}
	\caption{\textbf{Bayesian Optimization.}
		a)~Calculated loss function (negative of the $\Delta R_M$ amplitude) for scenario (2) in Fig. 5 of the main manuscript plotted within the range of hBN thicknesses explored by the optimizer. 
		b)~Optimization steps of BOSS with total number of acquisitions $N$ that equal to 10, 20, 50 and 100, demonstrating how the optimizer iteratively learns the loss function and estimates the global minimum.
	}
	\label{fig:Fig-S3}
\end{figure*}

\clearpage
\section{Transfer Matrix Method}
Throughout this work, the transfer matrix method (TMM) is used to calculate the reflectance of multilayer structures. In this approach, the electromagnetic field in each layer is expressed as a superposition of forward- and backward-propagating plane waves. Propagation through a layer is described by a propagation matrix that accounts for the accumulated phase, while interface matrices enforce the continuity conditions between adjacent media. Multiplying these $2\times2$ matrices in the order encountered by the incident wave yields the transfer matrix of the entire stack, from which the reflectance can be calculated directly. A detailed description of the method can be found in Ref.~\cite{byrnes2016multilayer}.

\clearpage
\section{Optimization parameters}
Here we provide further information about the machine learning-assisted stack optimization from Section VI of the main manuscript. 
Scenario (1) represents a typical scenario involving a CrSBr crystal on top of a Si-\ce{SiO2} substrate that does not require advanced optimization techniques. 
In an attempt to maximize the amplitude of $\Delta R_M$, we explore other structures commonly encountered in experiments, such as scenarios (2) and (3), in which the CrSBr crystal is encapsulated with hBN layers on each side. 
We optimize the thickness of these hBN layer. 

The optimized parameters are given  in Table I:
\begin{table}[h]
    \centering
    \begin{tabular}{c|c|c|c|c|c}
        \hline
        Scenario & \multicolumn{5}{c}{Material ( Thickness (nm) )}\\ \hline
         (1) & Si (Substrate) & \ce{SiO2} (285) & CrSBr (5)  &     &   \\ \hline
         (2) & Si (Substrate) & \ce{SiO2} (285) & hBN (158.3)  & CrSBr (5)  & hBN  (71) \\ \hline
         (3) & Au (Substrate) & hBN (108.1)  & CrSBr (5)  & hBN  (71.1)   &  \\ \hline
        \end{tabular}
    \caption{List of layer thicknesses after optimization.}
    \label{tab:my_label}
\end{table}

The list of refractive indices used during optimization is shown in Table II.
\begin{table}[h]
    \centering
    \begin{tabular}{c|c}
        \hline
          Material & Refractive index \\ \hline
        \ce{SiO2} & 1.48 \\ \hline
        Si & 3.77 \\ \hline
        hBN & 2.2 \\ \hline
        Au & 0.17 + 5.5i \\ \hline
    \end{tabular}
    \caption{List of refractive indices used in optimization.}
    \label{tab:my_label}
\end{table}

\clearpage

\bibliography{SI_Reflectance_paper}
\bibliographystyle{naturemag}